\documentclass[a4paper,10pt,twoside]{cpc-hepnp}

\usepackage{multicol}
\usepackage{graphicx}
\usepackage{booktabs}
\usepackage{amssymb,bm,mathrsfs,bbm,amscd}
\usepackage[tbtags]{amsmath}
\usepackage{lastpage}

\newcommand{\be}{\begin{equation}}
\newcommand{\ee}{\end{equation}}
\newcommand{\ba}{\begin{eqnarray}}
\newcommand{\ea}{\end{eqnarray}}

\newcommand{\De}{\Delta}

\newcommand{\im}{\text{Im}}

\begin{document}

\fancyhead[co]{\footnotesize E. Oset, dynamically generated resonances}

\footnotetext[0]{Received 14 March 2009}

\title{Dynamically generated resonances}

\author{%
      E. Oset$^{1}$\email{oset@ific.uv.es}%
\quad A. Ramos$^{2}$%
\quad S. Sarkar$^{3}$
\quad Bao Xi Sun$^{4}$
\quad M.J.Vicente Vacas$^{1}$
\quad P. Gonzalez$^{1}$
\quad J. Vijande$^{5}$\\
\quad D. Jido$^{6}$
\quad T. Sekihara$^{7}$
\quad A. Martinez-Torres$^{1}$
\quad K.Khemchandani$^{8}$
}
\maketitle

\address{%
1~Departamento de F\'{\i}sica Te\'orica and IFIC,
Centro Mixto Universidad de Valencia-CSIC,
Institutos de Investigaci\'on de Paterna, Aptdo. 22085, 46071 Valencia, Spain\\
2~Departament d'Estructura i
  Constituents de la Mat\`eria and Institut de Ci\`encies del Cosmos,
  Universitat de Barcelona, 08028 Barcelona, Spain\\
3~Variable Energy Cyclotron Centre, 1/AF, Bidhannagar, Kolkata 700064, India\\
4~Institute of Theoretical Physics, College of Applied Sciences,
Beijing University of Technology, Beijing 100124, China\\
5~Departamento de F\'{\i}sica Atomica Molecular y Nuclear and IFIC,
Centro Mixto Universidad de Valencia-CSIC,
Institutos de Investigaci\'on de Paterna, Aptdo. 22085, 46071 Valencia, Spain\\
6~Yukawa Institute for Theoretical Physics, Kyoto University, Kyoto 606-8502,
Japan\\
7~Department of Physics, Kyoto University, Kyoto 606-8502, Japan\\
8~Centro de F\'{\i}sica Computacional, Departamento de F\'{\i}sica,
Universidade de Coimbra, P-3004-516 Coimbra, Portugal\\
}

\begin{abstract}
In this talk I report on recent work related to the dynamical generation of
baryonic resonances, some made up from pseudoscalar meson-baryon, others from
vector meson-baryon and a third type from two meson-one baryon systems. We can
establish a correspondence with known baryonic resonances, reinforcing
conclusions previously drawn and bringing new light on the nature of some
baryonic resonances of higher mass.
\end{abstract}

\begin{keyword}
dynamically generated resonances, chiral dynamics, hidden gauge
formalism for vector meson interaction.
\end{keyword}

\begin{pacs}
13.75.Lb, 12.40.Vv, 12.40.Yx, 14.40.Cs
\end{pacs}

\begin{multicols}{2}

\section{Introduction}

The topic of dynamically generated resonances is bringing new light into the
interpretation of many mesonic and baryonic resonances \cite{review,puri}.
Considering the case of the baryons, the underlying physics can be depicted in
a schematic picture. It is well known that quarks can not be directly seen
because when we give energy to the nucleon to break it, the excitation energy
goes into creating meson pairs, mostly pions. This could equally happen even
if we just want to excite the nucleon, not to break it. Indeed, the two first
excitations of the nucleon correspond to
the Roper, $N^*(1440)$ and the $N^*(1535)$.  This means 500-600 MeV energy
excess over the nucleon mass, which in the quark model would correspond to
the excitation energy of one quark. However, the creation of one or two pions
costs less energy than this. So, why should not many baryon resonances
correspond to bound states or resonant states of a meson and a ground state of
the baryons, or even two mesons and one baryon? The possibility that this occurs
depends on whether the dynamics of the meson baryon interaction provides 
enough attraction to stabilize the system.
Fortunately there is an excellent theory to study these interactions at low
energies which is based on effective chiral Lagrangians that implement 
the chiral symmetry of the underlying
QCD \cite{Eck95,Be95}. In the present case we shall also report on recent results
for the interaction of vector mesons with baryons, which require new
Lagrangians. Here again one is lucky that such information is available in a
scheme which is an extension of the chiral Lagrangians to incorporate vector
mesons. The formulation starts by demanding invariance of these latter Lagrangians
under local gauge
transformations, that require the introduction of vector mesons
which obtain their mass through the mechanism of spontaneous symmetry breaking. This
scheme is known as the hidden gauge formalism which we shall use here too
\cite{hidden1,hidden2,hidden3}.

     The novel topics that I will report about are developments on
learning about the nature of resonances, which has been recently described in
 \cite{Hyodo:2008xr}, the recent work showing extra evidence for the existence
 of two $\Lambda(1405)$ states \cite{sekihara}, the generation of resonances from two mesons and
 a baryon \cite{alberto1,alberto2}, particularly one around 1920 MeV
 \cite{Jido:2008zz,KanadaEn'yo:2008wm,albertopheno} that could have been observed
 experimentally, and finally the novel work on the generation of resonances from
 the interaction  of vector mesons with baryons, either from the octet
 \cite{angelsvec} or the decuplet \cite{souravbao} of ground
 state baryons.

\section{Searching for the nature of resonances}

In this section I will report upon the work of \cite{Hyodo:2008xr}.
In this work one considers the
scattering of a pseudoscalar meson with mass $m$ from a target baryon with
mass $M_T$. The $s$-channel two-body unitarity condition for the amplitude
$T(\sqrt{s})$ can be expressed as
\begin{equation}
    \im T^{-1}(\sqrt{s}) = \frac{\rho(\sqrt{s})}{2} ,
    \label{eq:unitarity}
\end{equation}
where $\rho(\sqrt{s})=2M_T\bar{q}/(4\pi\sqrt{s})$ is the two-body phase
space of the scattering system with
$\bar{q}=\sqrt{[s-(M_T-m)^2][s-(M_T+m)^2]}/(2\sqrt{s})$. This is the
so-called elastic unitarity. Based on the $N/D$ method~\cite{nsd,ollerulf},
the general form of the scattering amplitude satisfying
Eq.~\eqref{eq:unitarity} is given by
\begin{align}
    T(\sqrt{s})
    =& \frac{1}{V^{-1}(\sqrt{s})-G(\sqrt{s})} ,
    \label{eq:TChU}
\end{align}
where $V(\sqrt{s})$ is a real function expressing the dynamical contributions
other than the $s$-channel unitarity and will be identified as the kernel
interaction. $G(\sqrt{s})$ is obtained by the once subtracted dispersion
relation with the phase-space function $\rho(\sqrt s)$. An analytical
expression for this $G(\sqrt{s})$ function can be found in \cite{ollerulf} which
shows explicitly the subtraction constant $a(\mu)$ of the dispersion
relation:
\begin{align}
    G(\sqrt{s})=&\frac{2M_T}{(4\pi)^{2}}
    \Bigl\{a(\mu)+\ln\frac{M_T^{2}}{\mu^{2}}
    +\frac{m^{2}-M_T^{2}+s}{2s}\ln\frac{m^{2}}{M_T^{2}}
    \nonumber\\
    &+\frac{\bar{q}}{\sqrt{s}}
    [\ln(s-(M_T^{2}-m^{2})+2\sqrt{s}\bar{q})
    \nonumber\\
    &
    +\ln(s+(M_T^{2}-m^{2})+2\sqrt{s}\bar{q})
    \nonumber\\
    &
    -\ln(-s+(M_T^{2}-m^{2})+2\sqrt{s}\bar{q})
    \nonumber\\
    &
    -\ln(-s-(M_T^{2}-m^{2})+2\sqrt{s}\bar{q})
    ]\Bigr\} .
    \label{eq:Gdim}
\end{align}

 One usually assumes $V$ to be given by the Weinberg-Tomozawa interaction and
 then one has
\begin{equation}
   T(\sqrt s) = \frac{1}{V^{-1}_{\text{WT}}(\sqrt s)
   - G(\sqrt s; a_{\rm pheno})}
   \label{eq:ampPheno} ,
\end{equation}
with the subtraction constant $a_{\text{pheno}}$ in the loop function $G$
being a free parameter to reproduce experimental data. This scheme can
describe various phenomena well, but the subtraction constant does not
always satisfy the natural renormalization condition  of \cite{ollerulf}, which
corresponds to having an equivalent $G$ function using a cut off of the order of
700-1000 MeV, the scale of the effective theory.

One can achieve an equivalent
scattering amplitude, using a different interaction kernel
$V_{\rm natural}$ as
\begin{equation}
   T(\sqrt s) = \frac{1}{V_{\rm natural}^{-1}(\sqrt s) -
   G(\sqrt s; a_{\rm natural})}
   \label{eq:ampNatural} .
\end{equation}
 The interaction kernel $V_{\rm natural}$ should be
modified from $V_{\text{WT}}$ in order to reproduce experimental
observables. Thus, equating the denominators of
Eqs.~\eqref{eq:ampPheno} and \eqref{eq:ampNatural}
one obtains \cite{Hyodo:2008xr}:
\begin{align}
    V_{\text{natural}}(\sqrt s)
    = &-\frac{C}{2f^2}(\sqrt{s}-M_T)+\frac{C}{2f^2}
    \frac{(\sqrt{s}-M_T)^2}{\sqrt{s}-M_{\text{eff}}}
    \label{eq:pole2}
\end{align}

   In Eq. (\ref{eq:pole2}) the first term represents the Weinberg-Tomozawa
    interaction, the seed to generate dynamically the resonances, and the
    second term would represent the contribution to account for a
    genuine part of the wave function.  The findings of \cite{Hyodo:2008xr}
    indicate that, while the $\Lambda(1405)$ is essentially a pure dynamically
    generated state, the $N^*(1535)$ demands also a genuine component, probably a
    three quark component.

     A very recent work in which a pole to account for a possible genuine
component of the $N^*(1535)$ is considered together with the driving
Weinberg-Tomozawa term, reinforces the leading role of the dynamically generated
$N^*(1535)$ component, once another pole to account for a genuine
$N^*(1650)$ (non pseudoscalar-baryon state for this purpose) is considered
\cite{mishanaka}.

\section{The $K^- d \to n\Lambda(1405)$ reaction and further evidence for the
existence of two $\Lambda(1405)$ states}

   In the chiral SU(3) framework for meson-baryon interaction one has the
interaction of one octet of mesons with the octet of baryons, which leads to a
singlet, a symmetric octet and an antisymmetric octet in which the interaction is
attractive, while it is repulsive in the other multiplets \cite{cola}. This leads to
two octets and a singlet of dynamically generated states and the two octets are
degenerate in the SU(3) limit when the masses of the mesons are made equal as
well as those of the baryons. As the physical masses are gradually restored, the
two octets split apart and one of them approaches the pole of the singlet at
energies around 1400 MeV, such that they overlap and the physical
$\Lambda(1405)$ is a superposition of the two resonances. Yet, there are some
differences between the two states: the one at 1395 MeV is wide and
couples strongly to $\pi \Sigma$, while that at 1420 MeV is narrow (around 30
MeV) and couples mostly to $\bar{K} N$. These
findings have been corroborated by all following chiral dynamical works on this
issue
\cite{carmina,carmenjuan,hyodo,Hyodo:2006kg,Borasoyweise,Borasoyulf,Oller:2006jw}.
Because of the different shape of the two states and the different coupling to
$\bar{K}N$ or  $\pi \Sigma$, one expects that the $\Lambda(1405)$ will show up
with different positions and widths in different experiments, depending on whether
it is produced by an initial $\bar{K}N$ or  $\pi \Sigma$ state.  The first
evidence for this was seen in the $K^- p \to \pi^0 \pi^0 \Sigma^0$ experiment
\cite{Prakhov:2004an} where a narrow peak was found around 1420 MeV. The
reaction was studied theoretically in \cite{Magas:2005vu}, where it was shown
that the resonance was indeed excited from the $\bar{K}N$ channel and this
was responsible for the  experimental shape as predicted in \cite{cola}.

   A further evidence for it has come from the analysis of the experiment of
\cite{Braun:1977wd} done recently in \cite{sekihara}. The reaction is
$K^- d \to n \Lambda(1405)$, but the peak of the resonance is seen clearly at
1420 MeV. It is curious to see how the $\Lambda(1405)$ can be made in a
$K^- p$ reaction when the resonance is below the $K^- p$ threshold. The answer is found
in \cite{sekihara}, where the reaction was studied taking into account single
and double scattering of the $K^-$, as depicted in fig. \ref{fig2}. The reaction
proceeds basically by double scattering: in a first scattering there is a
collision of the $K^-$ which gives energy to a neutron and brings the $K^-$ below
threshold to produce the $\Lambda(1405)$ in the second collision.  The
calculated cross section of \cite{sekihara} agrees well with experiment in
shape and size, leaving apart a bump around 1385 MeV that in \cite{sekihara}
is found to come from $\Sigma(1385)$ excitation, the inclusion of which does not
distort the shape of the $\Lambda(1405)$. Experiments in this line are planned
for J-PARC.

\begin{center}
\includegraphics[width=8.5cm]{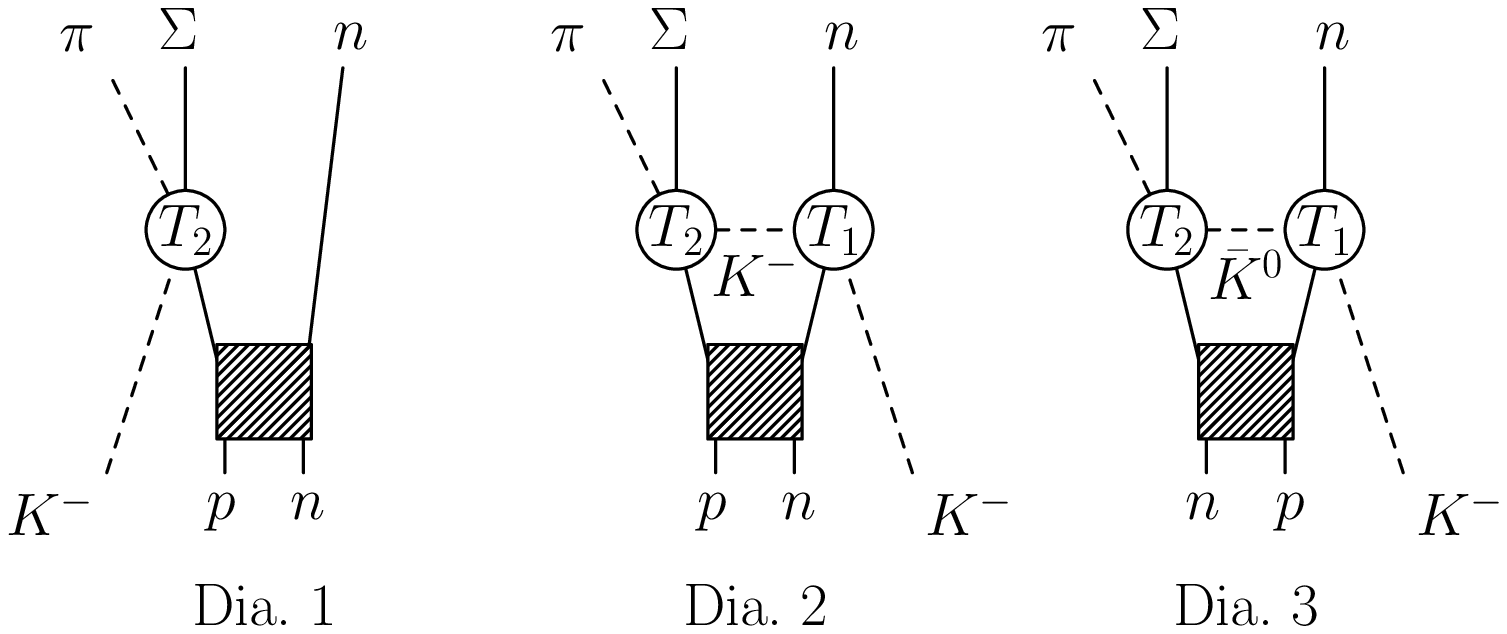}
\figcaption{Diagrams for the calculation of the $K^{-}d \to \pi\Sigma n$ reaction.
$T_{1}$ and $T_{2}$ denote the scattering amplitudes for $\bar KN \to \bar KN$
and $\bar K N \to \pi \Sigma$, respectively.  \label{fig2}}
\end{center}

\begin{center}
\includegraphics[width=7.5cm]{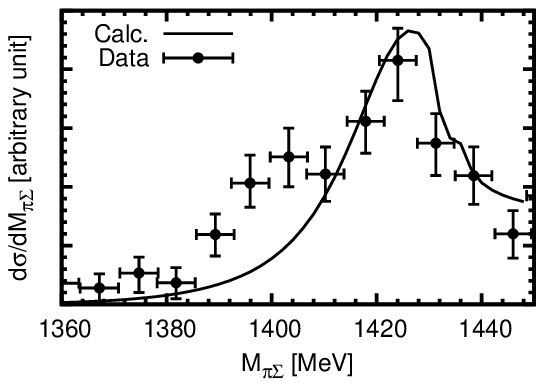}
\figcaption{\label{fig:IMspec} $\pi \Sigma$ invariant mass spectra of
$K^{-}d \rightarrow \pi^{+}\Sigma^{-}n$ in arbitrary units
at 800 MeV/c incident $K^{-}$ momentum.
The solid line denotes the present calculation.
The data are taken from the bubble chamber experiment at $K^{-}$ momenta
between 686
and 844 MeV/c, see text}.
\end{center}

\section{States of two mesons and a baryon}
 There are two specific talks on this issue in the Workshop
 \cite{albertotalk,jidotalk}. I
 will summarize a bit the important findings in this area by different groups.
In \cite{alberto1,alberto2} a formalism was develop to study Faddeev equations
of systems of two mesons and a stable baryon. The interaction of the pairs
was obtained from the chiral unitary approach, which proves quite successful to
give the scattering amplitudes of meson-meson and meson-baryon systems in the
region of energies of interest to us. The spectacular finding is that,
leaving apart the Roper resonance, whose structure is far more elaborate than
originally thought \cite{Krehl:1999km,Dillig:2004rh}, all the low lying
$J^P=1/2^+$ excited states are obtained as bound states or resonances of two
mesons and one baryon in coupled channels.

  It is rewarding to see that the idea is catching up and an independent
study, using variational methods found a bound state of $K \bar{K} N$, with
the $K \bar{K}$ being in the $a_0(980)$ state \cite{Jido:2008zz}.  The system
was studied a posteriori in \cite{albertopheno} and it was found to
appear at the same energy and the same configuration, although with a mixture
of $f_0(980) N$, see fig. \ref{threebody}. This state appears around 1920 MeV with $J^P=1/2^+$. In a
recent paper \cite{albertoulf} some arguments were given to associate this state
with the bump that one sees in the $\gamma p \to K^+ \Lambda$ reaction around
this energy, which is clearly visible in recent accurate experiments
\cite{Bradford:2005pt,Sumihama:2005er}. If this association was correct there
would be other  experimental consequences, as an enhanced strength of the
$\gamma p \to K^+ K^- p$ cross section close to threshold, as well as a shift
of strength close to the $K \bar{K}$ threshold in the invariant mass
distribution of the kaon pair.  This experiment is right now under study and
preliminary results corroborate our predictions \cite{nakanotalk}.

\begin{center}
\includegraphics[scale=0.8]{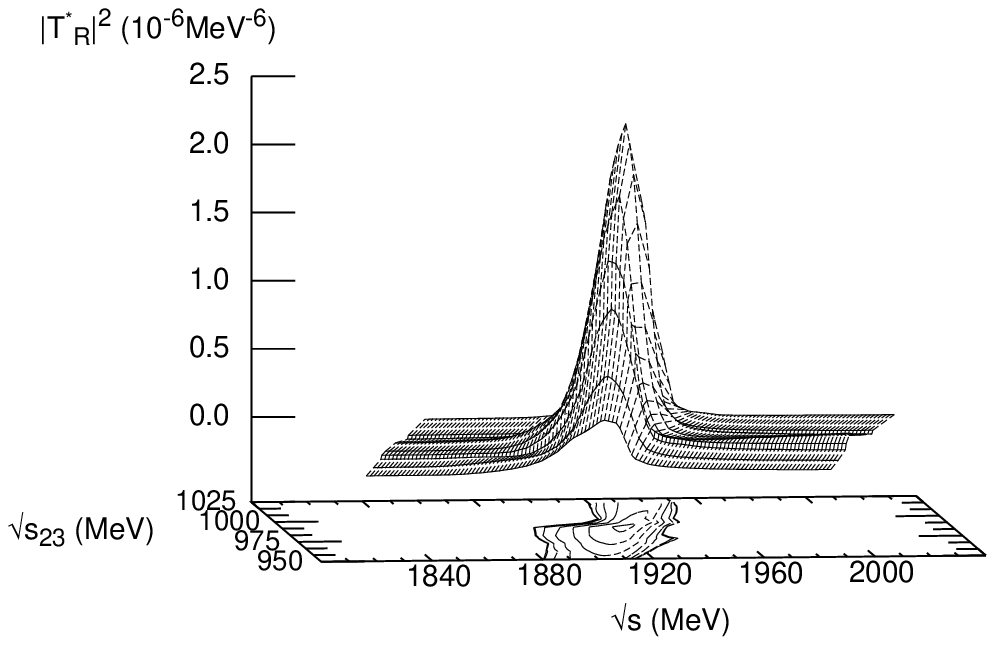}
\figcaption{\label{threebody}A possible $N^*(1910)$ in the $N K\bar{K}$ channels.}\label{N_1910}
\end{center}

\section{Resonances from the interaction of vector mesons with baryons}

  This is a very novel development since, as we shall see, some of the high mass
baryon resonances can be represented like bound states of vector mesons and
baryons, either from the octet of stable baryons or the decuplet.

\subsection{Formalism}

We follow the formalism of the hidden gauge interaction for vector mesons of
\cite{hidden1,hidden2,hidden3,ulfvec} (see also \cite{hidekoroca} for a practical set of Feynman rules).
The  Lagrangian involving the interaction of
vector mesons amongst themselves is given by
\begin{equation}
{\cal L}_{III}=-\frac{1}{4}\langle V_{\mu \nu}V^{\mu\nu}\rangle \ ,
\label{lVV}
\end{equation}
where the symbol $\langle \rangle$ stands for the trace in the $SU(3)$ space
and $V_{\mu\nu}$ is given by
\begin{equation}
V_{\mu\nu}=\partial_{\mu} V_\nu -\partial_\nu V_\mu -ig[V_\mu,V_\nu]\ ,
\label{Vmunu}
\end{equation}
where  $g$ is
\begin{equation}
g=\frac{M_V}{2f}\ ,
\label{g}
\end{equation}
with $f=93$ MeV the pion decay constant.  The magnitude $V_\mu$ is the $SU(3)$
matrix of the vectors of the octet of the $\rho$
\begin{equation}
V_\mu=\left(
\begin{array}{ccc}
\frac{\rho^0}{\sqrt{2}}+\frac{\omega}{\sqrt{2}}&\rho^+& K^{*+}\\
\rho^-& -\frac{\rho^0}{\sqrt{2}}+\frac{\omega}{\sqrt{2}}&K^{*0}\\
K^{*-}& \bar{K}^{*0}&\phi\\
\end{array}
\right)_\mu \ .
\label{Vmu}
\end{equation}

The lagrangian ${\cal L}_{III}$ gives rise to a contact term coming from
$[V_\mu,V_\nu][V_\mu,V_\nu]$
\begin{equation}
{\cal L}^{(c)}_{III}=\frac{g^2}{2}\langle V_\mu V_\nu V^\mu V^\nu-V_\nu V_\mu
V^\mu V^\nu\rangle\ ,
\label{lcont}
\end{equation}
as well as to a three
vector vertex which can be conveniently rewritten  as
\begin{eqnarray}
{\cal L}^{(3V)}_{III}
=ig\langle (V^\mu\partial_\nu V_\mu -\partial_\nu V_\mu
V^\mu) V^\nu\rangle
\label{l3Vsimp}\ .
\end{eqnarray}
In this case one finds an analogy to the coupling of vectors to
 pseudoscalars given in the same theory by
\begin{equation}
{\cal L}_{VPP}= -ig \langle [
P,\partial_{\nu}P]V^{\nu}\rangle \ ,
\label{lagrVpp}
\end{equation}
where $P$ is the SU(3) matrix of the pseudoscalar fields.

In a similar way, one obtains the Lagrangian for the coupling of vector mesons to
the baryon octet given by
\cite{Klingl:1997kf,Palomar:2002hk} \footnote{Correcting a misprint in
\cite{Klingl:1997kf}}
\begin{equation}
{\cal L}_{BBV} =
g\left( \langle \bar{B}\gamma_{\mu}[V^{\mu},B]\rangle +
\langle \bar{B}\gamma_{\mu}B \rangle \langle V^{\mu}\rangle \right)
\label{lagr82}
\end{equation}
where $B$ is now the SU(3) matrix of the baryon octet
\begin{equation}
B =
\left(
\begin{array}{ccc}
\frac{1}{\sqrt{2}} \Sigma^0 + \frac{1}{\sqrt{6}} \Lambda &
\Sigma^+ & p \\
\Sigma^- & - \frac{1}{\sqrt{2}} \Sigma^0 + \frac{1}{\sqrt{6}} \Lambda & n \\
\Xi^- & \Xi^0 & - \frac{2}{\sqrt{6}} \Lambda
\end{array} \ .
\right)
\end{equation}

With these ingredients we can construct the Feynman diagrams that lead to the $PB
\to PB$ and $VB \to VB$ interaction, by exchanging a vector meson between the
pseudoscalar or the vector meson and the baryon, as depicted in Fig.
\ref{fig:feyn}.

\begin{center}
\includegraphics[width=0.4\textwidth]{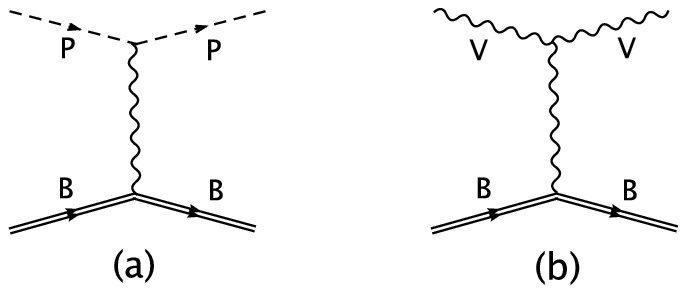}
\figcaption{\label{fig:feyn}Diagrams contributing to the pseudoscalar-baryon (a) or vector-
baryon (b) interaction via the exchange of a vector meson.}
\end{center}

From the diagram of Fig. \ref{fig:feyn}(a), and under the low energy approximation of
neglecting $q^2/M_V^2$ in the propagator of the exchanged vector, where $q$ is the
momentum transfer, one obtains the
same amplitudes as obtained from the ordinary chiral Lagrangian for
pseudoscalar-baryon octet interaction \cite{Eck95,Be95}, namely the Weinberg-Tomozawa
terms.  The approximation of neglecting the three momenta of the vectors implies
that $V^{\nu}$ in eq. (\ref{l3Vsimp}) corresponds to the exchanged vector and the analogy
with eq. (\ref{lagrVpp}) is more apparent. Note that $\epsilon_\mu \epsilon^\mu$ becomes
$-\vec{\epsilon}\,\vec{\epsilon }\,^\prime$ and the signs of the Lagrangians also agree.

   A small amendment is in order in the case of vector mesons, which
   is due to the mixing of $\omega_8$ and the singlet of SU(3), $\omega_1$, to give the
   physical states of the $\omega$ and the $\phi$ mesons:
   \begin{eqnarray}
   \omega=\sqrt{\frac{2}{3}} \omega_1 + \frac{1}{\sqrt 3} \omega_8 \nonumber \\
   \phi=\frac{1}{\sqrt 3} \omega_1 - \sqrt{\frac{2}{3}} \omega_8
   \label{eq:omephi}
   \end{eqnarray}
   Given the structure of Eq.~(\ref{eq:omephi}), the singlet state  which is accounted for
   by the V matrix, $diag(\omega_1,\omega_1,\omega_1)/\sqrt3$, does not provide
   any contribution to Eq.~(\ref{l3Vsimp}), in which case all one must do is to take the
   matrix elements known for the $PB$ interaction and, wherever $P$ corresponds to the
   $\eta_8$, the amplitude should be multiplied by the factor $1/\sqrt 3$
   to get the corresponding $\omega $ contribution, and by $-\sqrt{2/3}$ to get the
   corresponding $\phi$ contribution. Upon the approximation consistent with
   neglecting the three momentum versus the mass of the particles (in this
   case the baryon), we can just take the $\gamma^0$ component of
   Eq. (\ref{lagr82})  and
   then the transition potential corresponding to the diagram of \ref{fig:feyn}(b) is
   given by
   \begin{equation}
V_{i j}= - C_{i j} \, \frac{1}{4 f^2} \, \left( k^0 + k^\prime{}^0\right)
~\vec{\epsilon}\,\vec{\epsilon }\,^\prime
\label{kernel}
\end{equation}
   where $k^0, k^\prime{}^0$ are the energies of the incoming and outgoing vector meson.

    The $C_{ij}$ coefficients of eq. (\ref{kernel}) can be obtained directly from
    \cite{angels,bennhold,inoue}
    with the simple rules given above for the $\omega$ and the $\phi$ mesons, and
    substituting $\pi$ by $\rho$ and $K$ by $K^*$ in the matrix elements. They can be
    found in the appendix of \cite{angelsvec} where
    one can see that the
     cases with $(I,S)=(3/2,0)$, $(2,-1)$ and $(3/2,-2)$, the last two
     corresponding to exotic channels, have a repulsive interaction
and do not produce poles in the scattering matrices.  However, the sectors
$(I,S)=(1/2,0)$, $(0,-1)$, $(1,-1)$ and $(1/2,-2)$ are attractive and one finds
 bound states and resonances in these cases.

    The scattering matrix is obtained  solving the
    coupled channels Bethe Salpeter equation in the on shell factorization approach of
    \cite{angels,ollerulf}
   \begin{equation}
T = [1 - V \, G]^{-1}\, V
\label{eq:Bethe}
\end{equation}
with $G$ being the loop function of a vector meson and a baryon of
 eq. (\ref{eq:Gdim}). This function
is convoluted with the spectral function of the vector mesons to take into
account their width as done in \cite{nagahiro}.

 In this
case the factor $\vec{\epsilon}\,\vec{\epsilon }\,^\prime$, appearing in the potential $V$,
factorizes also in the $T$ matrix for the external vector mesons. This trivial
spin structure is responsible for having degenerate states with spin-parity
$1/2^-, 3/2^-$ for the interaction of vectors with the octet of baryons and
$1/2^-, 3/2^-, 5/2^-$ for the interaction of vectors with the decuplet
 of baryons.

  What we have done here for the interaction of vectors with the octet of
  baryons can be done for the interaction of vectors with the decuplet of
  baryons, and the interaction is obtained directly from that of the
  pseudoscalar-decuplet of baryons studied in
  \cite{kolodecu,Sarkar:2004jh}. The study of this interaction
  in \cite{vijande,souravbao} leads
  also to the generation of many resonances which are described below.

We search for poles in the scattering matrices in the second Riemann sheet, as
defined in previous works \cite{luisaxial}, basically changing $\bar{q}_l$ by to $-\bar{q}_l$ in the
analytical formula of the $G$ function, Eq. (\ref{eq:Gdim}),
for channels where Re$(\sqrt s)$ is above the threshold of the corresponding
channel. From the residues of the amplitudes at the poles one obtains the
couplings of the resonances to the different channels.  Alternatively, one can
obtain these couplings from the amplitudes in the real axis as follows.
Assuming these amplitudes to behave as
\begin{equation}
T_{ij}=\frac{g_i g_j}{\sqrt s -M_R + i \Gamma /2} \ ,
\end{equation}
where $M_R$ is the position of the maximum of $\mid T_{ii}\mid$, with $i$ being
the channel to which the resonance couples more strongly, and $\Gamma$ its
width at half-maximum,
one then finds
\begin{equation}
\mid g_i \mid ^2= \frac{\Gamma}{2} \sqrt {|T_{ii}|^2} \ .
\label{couprealaxis}
\end{equation}
Up to a global phase, this expression allows one to determine the value of
$g_i$, which we take to be real. The other couplings are then derived from
\begin{equation}
g_j = g_i \frac{T_{ij}(\sqrt{s}=M_R)}{T_{ii}(\sqrt{s}=M_R)} \ .
\label{couprealaxis2}
\end{equation}

This procedure to obtain the couplings from  $|T|^2$ in the real axis was used
in \cite{junko} where it was found that changes in the input parameters which lead to
moderate changes in the position and the width of the states affected the
couplings more smoothly.

\subsection{Results}

In table \ref{tab:pdg} we show a summary of the results obtained from the
interaction of vectors with the octet of baryons and the tentative
association to known states \cite{pdg}.

\end{multicols}

\begin{center}
      \renewcommand{\arraystretch}{1.5}
     \setlength{\tabcolsep}{0.2cm}
\begin{tabular}{c|c|cc|ccccc}\hline\hline
$I,\,S$&\multicolumn{3}{c|}{Theory} & \multicolumn{5}{c}{PDG data}\\
\hline
    \vspace*{-0.3cm}
    & pole position    & \multicolumn{2}{c|}{real axis} &  &  & &  &  \\
    &   & mass & width &name & $J^P$ & status & mass & width \\
    \hline
$1/2,0$ & --- & 1696  & 92  & $N(1650)$ & $1/2^-$ & $\star\star\star\star$ & 1645-1670
& 145-185\\
  &      &       &     & $N(1700)$ & $3/2^-$ & $\star\star\star$ &
	1650-1750 & 50-150\\
       & $1977 + {\rm i} 53$  & 1972  & 64  & $N(2080)$ & $3/2^-$ & $\star\star$ & $\approx 2080$
& 180-450 \\	
   &     &       &     & $N(2090)$ & $1/2^-$ & $\star$ &
 $\approx 2090$ & 100-400 \\
 \hline
$0,-1$ & $1784 + {\rm i} 4$ & 1783  & 9  & $\Lambda(1690)$ & $3/2^-$ & $\star\star\star\star$ &
1685-1695 & 50-70 \\
  &       &       &    & $\Lambda(1800)$ & $1/2^-$ & $\star\star\star$ &
1720-1850 & 200-400 \\
       & $1907 + {\rm i} 70$ & 1900  & 54  & $\Lambda(2000)$ & $?^?$ & $\star$ & $\approx 2000$
& 73-240\\
       & $2158 + {\rm i} 13$ & 2158  & 23  &  &  &  & & \\
       \hline
$1,-1$ &  ---  & 1830  & 42  & $\Sigma(1750)$ & $1/2^-$ & $\star\star\star$ &
1730-1800 & 60-160 \\
  & ---    & 1987  & 240  & $\Sigma(1940)$ & $3/2^-$ & $\star\star\star$ & 1900-1950
& 150-300\\
   &     &       &   & $\Sigma(2000)$ & $1/2^-$ & $\star$ &
$\approx 2000$ & 100-450 \\\hline
$1/2,-2$ & $2039 + {\rm i} 67$ & 2039  & 64  & $\Xi(1950)$ & $?^?$ & $\star\star\star$ &
$1950\pm15$ & $60\pm 20$ \\
         & $2083 + {\rm i} 31 $ &  2077     & 29  &  $\Xi(2120)$ & $?^?$ & $\star$ &
$\approx 2120$ & 25  \\
 \hline\hline
    \end{tabular}
\tabcaption{\label{tab:pdg}The properties of the 9 dynamically generated resonances and their possible PDG
counterparts.}
\end{center}

\begin{multicols}{2}

  For the $(I,S)=(1/2,0)$ $N^*$  states there is the $N^*(1700)$ with
 $J^P=3/2^-$, which could correspond to the state we find with the same quantum
 numbers around the same energy. We also find in the PDG the  $N^*(1650)$, which
 could be the near degenerate spin parter of the $N^*(1700)$ that we predict in
 the theory. It is interesting to recall that in the study of
 Ref.~\cite{mishajuelich} a pole is found around 1700 MeV,
with the largest coupling to $\rho N$ states.
Around 2000 MeV, where we find another $N^*$ resonance,
there are the states $N^*(2080)$ and $N^*(2090)$, with $J^P=3/2^-$ and
$J^P=1/2^-$ respectively, showing a good approximate spin degeneracy.

For the case $(I,S)=(0,-1)$ there is in the PDG one state, the $\Lambda(1800)$
with $J^P=1/2^-$, remarkably close to the energy were we find a $\Lambda$
state.  The state obtained around 1900 MeV could
correspond to the $\Lambda(2000)$ cataloged in the PDG with unknown spin and parity.

 The case of the $\Sigma $ states having $(I,S)=(1,-1)$ is rather interesting.
 The state
that we find around 1830 MeV, could be associated to the  $\Sigma(1750)$
with $J^P=1/2^-$. More interesting seems to be the case of the state obtained around
1990 MeV that could be related to two PDG candidates, again
nearly degenerate, the $\Sigma(1940)$ and the $\Sigma(2000)$, with spin and
parity  $J^P=3/2^-$ and $J^P=1/2^-$ respectively.

  Finally, for the case of the cascade resonances, $(I,S)=(1/2,-2)$, we find
  two states, one  around 2040 MeV and the other one around 2080 MeV. There are two cascade states in
  the PDG around this energy region with spin parity unknown, the
  $\Xi(1950)$ and the $\Xi(2120)$.  Although the experimental
  knowledge of this sector is relatively poor, a program is presently running at
  Jefferson Lab to improve on this situation \cite{Nefkens:2006bc}.

    The case of the vector interaction with the decuplet is
similar and we show the results in Table \ref{tab:pdg2}

\end{multicols}

\begin{center}
      \renewcommand{\arraystretch}{1.5}
     \setlength{\tabcolsep}{0.2cm}
\begin{tabular}{c|l|cc|lcclc}\hline\hline
$S,\,I$&\multicolumn{3}{c|}{Theory} & \multicolumn{5}{c}{PDG data}\\\hline
        & pole position &\multicolumn{2}{c|}{real axis} & name & $J^P$ & status & mass & width \\
        &               & mass & width & \\\hline
$0,1/2$ & $1850+i5$   & 1850  & 11  & $N(2090)$ & $1/2^-$ & $\star$ & 1880-2180 & 95-414\\
        &             &       &     & $N(2080)$ & $3/2^-$ & $\star\star$ & 1804-2081 & 180-450\\
        &       &  $2270(bump)$ &  & $N(2200)$ & $5/2^-$ & $\star\star$ & 1900-2228 & 130-400\\	
\hline
$0,3/2$ & $1972+i49$  & 1971  & 52  & $\De(1900)$ & $1/2^-$ & $\star\star$ & 1850-1950 & 140-240 \\	
	&             &       &     & $\De(1940)$ & $3/2^-$ & $\star$ & 1940-2057 & 198-460   \\
        &             &       &     & $\De(1930)$ & $5/2^-$ & $\star\star\star$ & 1900-2020  & 220-500   \\
	&             & $2200 (bump)$  &     & $\De(2150)$ & $1/2^-$ & $\star$ & 2050-2200  & 120-200  \\
\hline
$-1,0$  & $2052+i10$  & 2050  & 19  & $\Lambda(2000)$ & $?^?$ & $\star$  & 1935-2030 & 73-180\\
\hline
$-1,1$  & $1987+i1$   & 1985  & 10   & $\Sigma(1940)$ & $3/2^-$  & $\star\star\star$ &
1900-1950 & 150-300 \\
        & $2145+i58$  & 2144  & 57  & $\Sigma(2000)$ & $1/2^-$  & $\star$ & 1944-2004 &
	116-413\\
	& $2383+i73$  & 2370  & 99 & $\Sigma(2250)$ & $?^?$ & $\star\star\star$ & 2210-2280 &
	60-150\\
	&   &   &  & $\Sigma(2455)$ & $?^?$ & $\star\star$ & 2455$\pm$10 &
	100-140\\
\hline
$-2,1/2$ & $2214+i4$  & 2215  & 9  & $\Xi(2250)$ & $?^?$ & $\star\star$ & 2189-2295 & 30-130\\
	 & $2305+i66$ & 2308  & 66 & $\Xi(2370)$ & $?^?$ & $\star\star$ & 2356-2392 & 75-80 \\
         & $2522+i38$ & 2512  & 60  & $\Xi(2500)$ & $?^?$ & $\star$ & 2430-2505 & 59-150\\	 
\hline
$-3,1$   & $2449+i7$   & 2445 & 13  & $\Omega(2470)$   & $?^?$	 & $\star\star$ & 2474$\pm$12 & 72$\pm$33\\
 \hline\hline
 \end{tabular}
 \vspace{0.2cm}
   \tabcaption{\label{tab:pdg2}The properties of the 10 dynamically generated resonances and their possible PDG
counterparts. We also include the $N^*$ bump around 2270 MeV and the $\Delta^*$ bump around 2200 MeV. }
\end{center}

\begin{multicols}{2}

 We also can see that in many cases the experiment shows the near degeneracy
 predicted by the theory. Particularly, the case of the three $\Delta$
 resonances around 1920 MeV is very interesting. One observes a near
 degeneracy in the three spins $1/2^-, 3/2^-, 5/2^-$, as the theory predicts. It
 is also very instructive to recall that the case of the  $\Delta(5/2^-)$ is
 highly problematic in quark models since it has a  $3~h\omega$ excitation
 and comes out always with a very high mass \cite{vijande,pedro}.

  The association of states found to some reported in the PDG
  for the case of $\Lambda$, $\Sigma $ and $\Xi$ states looks also equally
  appealing as one can see from the table.

  In summary, the study of the interaction of mesons in the vector octet of the $\rho$
  with baryons of the octet of the proton and the decuplet of the $\Delta$ within the hidden gauge formalism of
  vector mesons, using a unitary framework in coupled channels, has lead to 
  a rich structure of excited baryons. Many of the states predicted
  by the theory can be associated to known states in the PDG, thus providing a very different
  explanation for the nature of these states than the one given by quark models as simple $3q$
  states. One of the particular predictions of the theory is that, within the
  approximations done, one obtains degenerate pairs of particles in
  $J^P= 1/2^-,3/2^-$ for the case of the interaction of vectors with the baryons
  of the octet and degenerate trios $J^P= 1/2^-,3/2^-,5/2^-$ for the case of
  the interaction of vectors with the baryons of the
  decuplet.
    This behavior seems well reproduced by many of the
  existing data, but in some cases the spin partners do not show up in the PDG.
  The reasonable results reported here produced by the hidden gauge approach
   should give a stimulus to
  search experimentally for the missing spin partners of the already observed
  states, as well as possible new ones.

\end{multicols}

\acknowledgments{This work is partly supported by the EU contract No. MRTN-CT-2006-035482
(FLAVIAnet), by the contracts FIS2006-03438 FIS2008-01661 from MICINN
(Spain) and by the Ge\-ne\-ra\-li\-tat de Catalunya contract 2005SGR-00343. We
acknowledge the support of the European Community-Research Infrastructure
Integrating Activity ``Study of Strongly Interacting Matter'' (HadronPhysics2,
Grant Agreement n. 227431) under the Seventh Framework Programme of EU.}

\vspace{-2mm}
\centerline{\rule{80mm}{0.1pt}}
\vspace{2mm}

\begin{multicols}{2}

\end{multicols}

\clearpage

\end{document}